\documentclass[universe,article,submit,pdftex,moreauthors]{Definitions/mdpi}
\usepackage{amsmath,amssymb,enumitem} 

\usepackage{bm}
\newcommand{\be}{\begin{equation}}
\newcommand{\ee}{\end{equation}}

\preto{\abstractkeywords}{\nolinenumbers}

\firstpage{1}
\pubvolume{1}
\issuenum{1}
\articlenumber{0}
\pubyear{2024}
\copyrightyear{2024}
\datereceived{ }
\daterevised{ } 
\dateaccepted{ }
\datepublished{ }
\hreflink{https://doi.org/} 

\Title{Gravitational Wave and Quantum Graviton Interferometer Arm Detection of Gravitons}
\TitleCitation{Gravitational Wave and Detection of Gravitons}

\Author{John W. Moffat $^{1}$\orcidA{}}

\AuthorNames{John W. Moffat}

\AuthorCitation{Moffat, J. W.}

\address{%
$^{1}$ \quad Perimeter Institute for Theoretical Physics, Waterloo, Ontario N2L 2Y5, Canada\\
and\\
Department of Physics and Astronomy, University of Waterloo, Waterloo,\\
Ontario N2L 3G1, Canada}

\corres{Correspondence: jmoffat@perimeterinstitute.ca}

\abstract{This paper explores the quantum and classical descriptions of gravitational wave detection in interferometers like LIGO. We demonstrate that a graviton scattering and quantum optics model succeeds in explaining the observed arm displacements, while the classical gravitational wave approach and a quantum graviton energy method also successfully predict the correct results. We provide a detailed analysis of why the quantum graviton energy approach succeeds, highlighting the importance of collective behavior and the quantum-classical correspondence in gravitational wave physics.
Our findings contribute to the ongoing discussion about the quantum nature of gravity and its observable effects in macroscopic physics.
}

\begin{document}
\maketitle


\section{Introduction}

An important problem in modern physics is to understand the role of quantum physics in gravity. A quantum gravity based on the quantization methods of effective quantum field theory assumes that the quantum particle is a graviton. The problem that has beset this approach has been the lack of experimental data allowing for a test of any quantum gravity predictions. At the basic level the length and time scales associated with Newton's gravitational constant $G$ are $l_p=\sqrt{G\hbar/c^3}\approx 10^{-35} m$ and $t_p=l_p/c\approx 10^{-44}s$. In effective field theory (EFT) and quantum mechanics, these scales are far beyond the reach of observational experiments, when single or a few gravitons are describing calculations of graviton scattering amplitudes and cross sections~\cite{Boughn,Dyson}. 

In ref.~\cite{Moffat1}, an attempt was made to replicate the stochastic Brownian motion detection of gravitons by proposing experiments to detect the stochastic spatial displacement 
$\Delta x$ of a test mass by a large number of collective coherent gravitons. 

Several proposals have been made for experiments to detect gravitons without resorting to a physically unrealistic large mass detector or accelerator~\cite{Pikovski1,Pikovski2,Parikh1,Parikh2,Parikh3,Cho,Kanno,Haba1,Haba2,Zurek1,Fuentes,HowlFuentes,Tabletop,Zurek2,Lee1,Lee2,Badurina,Bub,Carney,Carney24,Guerreiro,Abrahao}. 

The detection of gravitational waves by the Laser Interferometer Gravitational-Wave Observatories LIGO-Virgo-Kagra~\cite{Abbott} marked a milestone in physics, confirming a key prediction of Einstein's general relativity. These detections, resulting from cataclysmic events such as black hole mergers, involve enormous energies on the order of $10^{47}$ J, yet produce incredibly small measurable displacements in the LIGO interferometer arms, on the scale of $10^{-18}$~m.

This paper addresses a fundamental question in the quantum-classical correspondence of gravitational waves: How can we reconcile the quantum nature of gravity, described by gravitons, with the classical wave behavior observed in detectors? We explore three approaches to this problem: (1) A graviton scattering and quantum optics model, (2) The classical gravitational wave description, (3) A quantum graviton energy approach.

We demonstrate that all three approaches succeed in explaining the measurement of the LIGO-Virgo-Kagra arms displacement 
$\Delta L\sim 10^{-18}$~m. They provide valuable insights into the nature of quantum-classical correspondence in gravitational physics.

Our analysis sheds light on the collective behavior of gravitons, the importance of coherence in gravitational wave detection, and the fundamental ways in which gravitational waves interact with matter. These findings contribute to our understanding of quantum gravity and its observable effects in macroscopic systems.

\section{Classical gravitational Waves and gravitons}

Let us consider three approaches that can shed light on reconciling classical gravitational wave experiments with the quantum nature of gravity.

A) The classical gravitational energy approach.
The weak gravitational field is described by
\be
g_{\mu\nu}=\eta_{\mu\nu}+\kappa h_{\mu\nu},
\ee
where $\eta_{\mu\nu}$ is the Minkowski metric tensor and $\kappa=\sqrt{8\pi G/c^4}$. The coupling between matter and the gravitational field is proportional to the stress-energy momentum tensor $T_{\mu\nu}$ and the gravitational field perturbation $h_{\mu\nu}$. The displacement of the interferometer arms depends on the strength of this coupling. 

In the LIGO-Virgo-Kagra gravitational wave experiment, the freely falling test masses influenced by the gravitational tidal forces are primarily the mirrors at the ends of the interferometer arms. When a gravitational wave passes through the detector, it causes a differential change in the proper distance between the mirrors at the ends of the perpendicular arms. This differential change is what the interferometer measures. The geodesic deviation equation describes the relative acceleration of the nearby geodesics describing the motion of the test masses (i,j = 1,2,3):
\be
\frac{d^2\xi^i}{dt^2} = -R^i_{0j0}\xi^i,
\ee
where the $R^i_{0j0}$ are the Rieman tensor components and 
$\xi^i$ denotes the separation vector between the central beam splitter and one end of the arms. The Rieman tensor components are related to the strain:
\be
R^i_{0j0}\sim \frac{1}{2}\frac{d^2h_{ij}}{dt^2}.
\ee

The energy flux of a gravitational wave is given by
\be
F = \frac{c^3}{16\pi G}\langle{\dot h}_{ij}{\dot h}^{ij}\rangle,
\ee
where $\langle ...\rangle$ denotes the time average and 
${\dot h}_{ij} = dh_{ij}/dt$.

For a gravitational wave with strain amplitude $h_0$ and angular frequency $\omega$, the sinusoidal strain absorbed by the detector can be expressed as
\be
h(t) = h_0\sin(\omega t).
\ee
The time derivative of the strain is
\be
\dot h=\omega h_0\cos(\omega t),
\ee
and the time-averaged square of the derivative is given by
\be
\langle{\dot h}^2\rangle = \frac{1}{2}(\omega h_0)^2.
\ee

We obtain for the energy flux:
\be
F=\frac{c^3}{32\pi G}(2\pi f h_0)^2.
\ee
The total energy is the flux $F$ integrated over the surface sphere area $A = 4\pi r^2$, where $r$ is the distance to the merging black holes source of gravitational waves and $\Delta t$ is the time elapsed:
\be
E_{GW} = \frac{c^3}{32\pi G}(2\pi f h_0)^2 4\pi r^2\Delta t.
\ee
The time $\Delta t$ elapsed refers to the duration of the black hole merger as observed by the detectors. We have assumed that all gravitons have the same frequency, which together with the strain $h_0$ and the amplitude remain constant during the 
time $\Delta t$.

The gravitational wave energy emitted by the merger of two black holes is of order $3M_\odot c^2\sim 10^{47}$~J, while for the gravitational wave frequency $f\sim 200$~Hz the energy of a single graviton is $E_g = h f\sim 1.33\times 10^{-31}$~J, where $h$ is Planck's constant $h = 6.626\times 10^{-34}$~J$\cdot$s. The number of gravitons in the emitted gravitational wave is $E_{GW}/E_g = N_g\sim 10^{78}$. 

Let us consider the graviton scattering approach to solving the displacement of the interferometer arms. To determine the strain amplitude $h_0 =\Delta L/L$, the subtended angle 
$\theta$ of the beam of gravitons can be approximated by the ratio of the detector radius $r=1$~m to the distance of the black hole merger $d=10^{25}$~m:
\be
\theta\sim \frac{r}{d}\sim 10^{-25}~{\rm rad}.
\ee
This is an extremely small angle, which is expected given the vast distance to the source.

To calculate the number of gravitons penetrating the detector, we need to compare the detector area $A_{\rm det}=\pi r^2$, with $r\sim 1$~m, to the total area over which the gravitons are spread at the distance of the detector. The total area at the detector distance is given by
\be
A_{\rm tot}=4\pi d^2\sim 10^{51} m^2.
\ee
The fraction of gravitons hitting the detector is given by
\be
f = \frac{A_{\rm det}}{A_{\rm tot}}\sim 10^{-51}.
\ee
This tiny fraction $f$ makes it extremely unlikely for a single graviton or even a few gravitons to be detected.

Assuming all gravitons, $N_g\sim 10^{78}$, are emitted towards Earth, the number of gravitons hitting the detector is:
\be
N_{\rm gdet}=f\times N_g\sim 10^{27}.
\ee
We need to consider how quantum behavior of gravitons could potentially lead to the observed strain. This is a complex and speculative area, as we do not have a complete theory of quantum gravity. However, we can explore one possible approach that might contribute to a quantum explanation of the observed strain.

In quantum optics, coherent states are quantum states of the electromagnetic field that behave most like classical waves. An analogous concept might apply to gravitons. The gravitational wave could be described as a coherent state of gravitons, which would behave more like a classical wave than individual particles.

The amplitude of a coherent state is related to the average number of quanta. For a coherent state $|a\rangle$, the amplitude $a$ is related to the average number of quanta $\langle n\rangle$ by:
\be
|a|^2 = \langle n\rangle.
\ee
If we relate this to the strain $h_0$, we can propose the coherent state formula:
\be
h_0\sim \frac{a}{d}l_p,
\ee
where $d$ is the distance to the source, and 
$l_p$ is the Planck length. We obtain
\be
h_0\sim \frac{\sqrt{\langle n\rangle}}{d} l_p.
\ee
For $\langle n\rangle\sim 10^{78}$, $d=10^{25}$~ m and $l_p=1.62\times 10^{-35}$~m, we obtain 
\be
h_0\sim 10^{-21}.
\ee

This result is interesting, because the number of gravitons, $N_g\sim 10^{78}$, is very close to the one we calculated earlier based on the total energy of the gravitational wave black hole merger event. This suggests that if we consider gravitational waves as highly coherent states of gravitons, the effective number of gravitons in the state that produces the observed strain is similar to the total number of gravitons estimated from the energy of the black hole merger event.

This alignment between the classical energy calculation and the quantum coherent state model is intriguing. It suggests that a quantum description based on coherent states of gravitons could potentially be consistent with both the total energy of the gravitational wave event and the observed strain $h_0\sim 10^{-21}$ at Earth. This suggests that if gravitational waves are highly coherent states of gravitons, the effective number of gravitons $N_g\sim \langle n\rangle$ in the coherent state reaching Earth will be close to number calculated from the total energy $E_{GW}$ divided by the energy $E_g$ per graviton.

Let us consider the quantum graviton energy approach. The energy of a single graviton is $E_g = hf$, and the total quantum energy of $N_g$ gravitons is given by
\be
E_{GW} = N_ghf.
\ee
We equate this with the classical gravitational energy expression:
\be
\frac{c^3}{32\pi G}(2\pi f h_0)^2 4\pi r^2\Delta t
\sim N_ghf.
\ee
Solving for the strain $h_0$, we obtain
\be
h_0\sim\biggl(\frac{N_ghfG}{c^3\pi^2f^2r^2\Delta t}\biggr)^{1/2}.
\ee
For typical LIGO parameters: $f\sim 200 Hz,\Delta t\sim 0.1 s$, $d\sim 10^{25} m$ and $N_g\sim 10^{78}$, we predict the strain:
\be
h_0\sim 10^{-21}. 
\ee
This result is consistent with the LIGO measured strain $\Delta L/L\sim 10^{-21}$.

The quantum graviton energy approach, which bridges quantum and classical physics, correctly predicts the observed LIGO arms displacement $\Delta L\sim 10^{-18} m$ and the measured gravitational wave strain $h=\Delta L/L\sim 10^{-21}$.

\section{Discussion}

Let us explore why the quantum graviton energy approach succeeds: In the quantum graviton energy approach, we consider the total energy of the gravitational wave as composed of many gravitons:
\be
E_{GW} = N_gE_g,
\ee
where $E_g=hf$ is the energy of a single graviton. This total energy is then related to the classical strain through the equation we derived earlier:
\be
E_{GW} = \frac{c^3}{32\pi G}(2\pi fh_0)^2 4\pi r^2\Delta t.
\ee
The key point here is that this approach does not rely on individual graviton-matter interactions. Instead, it relates the collective energy of all gravitons to the classical wave properties.

The quantum graviton energy approach succeeds for several reasons:

a) It treats the gravitons as a collective ensemble, preserving their coherent nature.

b) It uses the fundamental relationship between energy and strain from general relativity, which encapsulates the correct physics of how gravitational waves interact with matter.

c) This approach respects the wave-particle duality of quantum mechanics. The gravitons are treated as particles in terms of their individual energies, but their collective effect is described as a wave.

d) This method effectively demonstrates the correspondence principle, showing how the many gravitons quantum description leads to the classical wave strain result.

The quantum graviton energy approach succeeds by bridging the quantum and classical descriptions. It starts with gravitons as the quantum entities, but relates their collective energy to classical wave properties, which are then used to calculate the detector response.

In essence, the quantum graviton energy approach provides a way to connect the graviton quantum nature of gravity with its classical gravitational wave manifestation in a manner that correctly predicts observable effects, demonstrating the deep connection between quantum and classical descriptions of gravity.

\section{Conclusions}

We have explored the intriguing interplay between quantum and classical descriptions of gravitational waves in the context of interferometric detection. Our analysis has led to several important conclusions. The graviton scattering and quantum optics models,  successfully explain the observed arm displacements in gravitational wave detectors like LIGO. This highlights the importance of treating gravitons as localized particles interacting through direct collisions with matter.

The classical gravitational wave approach, based on general relativity, also successfully predicts the correct arm displacements. This success underscores the power and accuracy of Einstein's theory in describing macroscopic gravitational phenomena.

Most significantly, the quantum graviton energy approach, which bridges quantum and classical concepts, correctly predicts the observed interferometer arm displacement $\Delta L/L\sim 10^{-21}$. This success is particularly noteworthy as it demonstrates a viable path for reconciling quantum and classical aspects of gravity in observable phenomena.

The success of both the graviton scattering and the quantum graviton energy approaches provide several key insights. It emphasizes the importance of considering the collective, coherent behavior of gravitons rather than their individual interactions. It illustrates how the transition from quantum to classical behavior can be understood through the lens of collective quantum properties. It demonstrates the deep connection between the graviton quantum nature of gravity and its classical gravitational wave manifestation.

These findings have broader implications for our understanding of quantum gravity and the quantum-classical correspondence in gravitational physics. They suggest that while individual quantum gravitational effects may be too weak to detect, their collective behavior manifests in measurable classical phenomena.

Future work could explore further implications of this quantum-classical bridge in gravitational wave physics. Potential avenues include investigating how this approach might inform quantum gravity theories, exploring possible quantum signatures in gravitational wave signals, and examining how this understanding might be applied to other astrophysical phenomena.

In conclusion, our analysis not only resolves the relation between quantum and classical descriptions of gravitational wave detection, but also provides a framework for understanding how quantum gravitational effects can manifest in macroscopic systems. This work contributes to the ongoing dialogue about the nature of quantum gravity and its observable consequences, paving the way for further theoretical and experimental investigations.

\section*{Acknowledgments}

I thank Viktor Toth and Martin Green for helpful discussions. Research at the Perimeter Institute for Theoretical Physics is supported by the Government of Canada through industry Canada and by the Province of Ontario through the Ministry of Research and Innovation (MRI).

\end{document}